# Experimental observation of the quantum anomalous Hall effect in a magnetic topological insulator

Cui-Zu Chang,[1,2]† Jinsong Zhang,[1]† Xiao Feng,[1,2]† Jie Shen,[2]† Zuocheng Zhang,[1] Minghua Guo,[1] Kang Li,[2] Yunbo Ou,[2] Pang Wei,[2] Li-Li Wang,[2] Zhong-Qing Ji,[2] Yang Feng,[1] Shuaihua Ji,[1] Xi Chen,[1] Jinfeng Jia,[1] Xi Dai,[2] Zhong Fang,[2] Shou-Cheng Zhang,[3] Ke He,[2]* Yayu Wang,[1]* Li Lu,[2] Xu-Cun Ma,[2] Qi-Kun Xue[1]*

[1]State Key Laboratory of Low-Dimensional Quantum Physics, Department of Physics, Tsinghua University, Beijing 100084, China

[2]Beijing National Laboratory for Condensed Matter Physics, Institute of Physics, The Chinese Academy of Sciences, Beijing 100190, China

[3]Department of Physics, Stanford University, Stanford, California 94305-4045 (USA)

* To whom correspondence should be addressed. E-mail:　qkxue@mail.tsinghua.edu.cn (Q.K.X.); kehe@iphy.ac.cn (K.H.); yayuwang@tsinghua.edu.cn (Y.W.)

† These authors contributed equally to this work.

**Abstract**: The quantized version of the anomalous Hall effect has been predicted to occur in magnetic topological insulators, but the experimental realization has been challenging. Here, we report the observation of the quantum anomalous Hall (QAH) effect in thin films of Cr-doped $(Bi,Sb)_2Te_3$, a magnetic topological insulator. At zero magnetic field, the gate-tuned anomalous Hall resistance reaches the predicted quantized value of $h/e^2$, accompanied by a considerable drop of the longitudinal resistance. Under a strong magnetic field, the longitudinal resistance vanishes whereas the Hall resistance remains at the quantized value. The realization of the QAH effect may lead to the development of low-power-consumption electronics.

Quantum Hall effect (QHE), a quantized version of the Hall effect (*1*), was observed in two-dimensional (2D) electron systems more than 30 years ago (*2,3*). In QHE the Hall resistance, which is the voltage across the transverse direction of a conductor divided by the longitudinal current, is quantized into plateaus of height $h/\nu e^2$, with $h$ being Planck's constant, $e$ the electron's charge, and $\nu$ an integer (*2*) or a certain fraction (*3*). In these systems, the QHE is a consequence of the formation of well-defined Landau levels, and thus only possible in high mobility samples and strong external magnetic fields. However, there have been numerous proposals to realize QHE without applying any magnetic field (*4-11*). Among these proposals, using the thin film of a magnetic topological insulator (TI) (*6-9,11*), a new class of quantum matter discovered recently (*12,13*), is one of the most promising routes.

Magnetic field induced Landau quantization drives a 2D electron system into an insulating phase that is topologically different from the vacuum (*14,15*); as a consequence, dissipationless states appear at sample edges. The topologically non-trivial electronic structure can also occur in certain 2D insulators with time reversal symmetry (TRS) broken by current loops (*4*) or by magnetic ordering (*6*), requiring neither Landau levels nor external magnetic field. This type of QHE induced by spontaneous magnetization is considered the quantized version of the conventional (non-quantized) anomalous Hall effect (AHE) discovered in 1881 (*16*). The quantized Hall conductance is directly given by a topological characteristic of the band structure called the first Chern number. Such insulators are called Chern insulators.

One way to realize a Chern insulator is to start from a time-reversal-invariant TI. These materials, whose topological properties are induced by spin-orbit coupling, were experimentally realized soon after the theoretical predictions in both 2D and 3D systems (*12,13*). Breaking the TRS of a suitable TI (*17*) by introducing ferromagnetism can naturally lead to the QAH effect (*6-9,11*). By tuning the Fermi level of the sample around the magnetically induced energy gap in the density of states, one is expected to observe a plateau of Hall conductance ($\sigma_{xy}$) of $e^2/h$ and a vanishing longitudinal conductance ($\sigma_{xx}$) even at zero magnetic field (Fig. 14 of (*7*) and Fig. 1, A and B).

The QAH effect has been predicted to occur by Mn doping of the 2D TI realized in HgTe quantum wells (8); however, an external magnetic field might still be required to align the Mn moments in order to realize the QAH effect (18). As proposed in (9), due to the van Vleck mechanism doping the $Bi_2Te_3$ family TIs with isovalent 3d magnetic ions can lead to a ferromagnetic insulator ground state, and for thin film systems, this will further induce QAH effect if the magnetic exchange field is perpendicular to the plane and overcomes the semiconductor gap. Here we investigate thin films of $Cr_{0.15}(Bi_{0.1}Sb_{0.9})_{1.85}Te_3$ (19, 20) with a thickness of 5 quintuple layers (QLs), which are grown on dielectric $SrTiO_3$ (111) substrates by molecular beam epitaxy (MBE) (20, 21, Fig. S1). With this composition, the film is nearly charge neutral so that the chemical potential can be fine-tuned to the electron- or hole-conductive regime by a positive or negative gate voltage, respectively, applied on the backside of the $SrTiO_3$ substrate (20). The films are manually cut into a Hall bar configuration (Fig. 1C) for transport measurements. Varying the width (from 50 μm to 200 μm) and the aspect ratio (from 1:1 to 2:1) of the Hall bar does not influence the result. Figure 1D displays a series of measurements, taken at different temperatures, of the Hall resistance ($\rho_{yx}$) of the sample in Fig. 1C, as a function of the magnetic field ($\mu_0 H$). At high temperatures, $\rho_{yx}$ exhibits linear magnetic field dependence due to the ordinary Hall effect (OHE). The film mobility is ~ 760 $cm^2$/(Vs), as estimated from the measured longitudinal sheet resistance ($\rho_{xx}$) and the carrier density determined from the OHE. The value is much enhanced compared with the samples in our previous study (20), but still much lower than that necessary for QHE (2,3). With decreasing temperature, $\rho_{yx}$ develops a hysteresis loop characteristic of the AHE, induced by the ferromagnetic order in the film (22). The square-shaped loop with large coercivity ($H_c$ = 970 Oersted at 1.5 K) indicates a long-range ferromagnetic order with out-of-plane magnetic anisotropy. The Curie temperature is estimated to be ~ 15 K (Fig. 1D, inset) from the temperature dependence of the zero field $\rho_{yx}$ that reflects spontaneous magnetization of the film.

Figure 2A and 2C show the magnetic field dependence of $\rho_{yx}$ and $\rho_{xx}$, respectively, measured at $T$ = 30 mK at different bottom-gate voltages ($V_g$s). The shape and coercivity of the $\rho_{yx}$ hysteresis loops (Fig. 2A) vary little with $V_g$, thanks to the robust ferromagnetism probably mediated by the van Vleck mechanism (9, 20). In the magnetized states, $\rho_{yx}$ is nearly independent of the magnetic field, suggesting perfect ferromagnetic ordering and charge neutrality of the sample. On the other hand, the AH resistance (height of the loops) changes dramatically with $V_g$, with a maximum value of $h/e^2$ around $V_g$ = -1.5 V. The magnetoresistance (MR) curves (Fig. 2C) exhibit the typical shape for a ferromagnetic material: two sharp symmetric peaks at the coercive fields.

The $V_g$ dependences of $\rho_{yx}$ and $\rho_{xx}$ at zero field (labeled $\rho_{yx}(0)$ and $\rho_{xx}(0)$, respectively) are plotted in Fig. 2B. The most important observation here is that the zero field Hall resistance exhibits a distinct plateau with the quantized value $h/e^2$, which is centered around the gate voltage $V_g$ = -1.5 V. This observation constitutes the discovery of the QAH effect. According to the OHE measurements, the maximum of $\rho_{yx}$ is always located at the charge neutral point (referred to as $V_g^0$ hereafter) (20, 21). Accompanying the quantization in $\rho_{yx}$, the longitudinal resistance $\rho_{xx}(0)$ exhibits a sharp dip down to 0.098 $h/e^2$. The $\rho_{yx}(0)/\rho_{xx}(0)$ ratio corresponds to a Hall angle of 84.4°. For comparison with theory, we transform $\rho_{yx}(0)$ and $\rho_{xx}(0)$ into sheet conductance via the relations: $\sigma_{xy} = \rho_{yx}/(\rho_{yx}^2+\rho_{xx}^2)$ and $\sigma_{xx} = \rho_{xx}/(\rho_{yx}^2+\rho_{xx}^2)$, and plot them in Fig. 2D. Around $V_g^0$, $\sigma_{xy}(0)$ has a notable plateau at 0.987 $e^2/h$, whereas $\sigma_{xx}(0)$ has a dip down to 0.096 $e^2/h$, similar to the behavior of the corresponding resistances.

In addition to the observation of the QAH effect, the MR ratio (($\rho_{xx}(H_c)-\rho_{xx}(0))/\rho_{xx}(0)$) is dramatically enhanced at $V_g^0$ to a surprisingly large value of 2251% (Fig. 1C and Fig. S3). The huge MR can also be understood in terms of the QAH phenomenology. In the magnetized QAH state, the existence of dissipationless edge state leads to a nearly vanishing $\rho_{xx}$ (21). At the coercive field, the magnetization reversal of a QAH system leads to a quantum phase transition between two QH states (7) via a highly dissipative phase with a large $\rho_{xx}$, though the exact process may be complex (23). The huge MR thus reflects the distinct difference in transport properties between an ordinary insulator and a QAH insulator.

For a QH system (2,3), when the Fermi level lies in the gap between Landau levels, $\sigma_{xy}$ reaches a plateau at $\nu e^2/h$ and $\sigma_{xx}$ drops to zero. If the system contains non-localized dissipative conduction channels, $\sigma_{xx}$ has a non-zero value, whereas $\sigma_{xy}$ deviates slightly from the quantized plateau (*24*). For a QAH system, only one $\sigma_{xy}$ plateau of $e^2/h$ appears at zero field when the Fermi level is around the magnetically induced gap (Fig. 1B). The observations of $\sigma_{xy}(0) = e^2/h$ plateau and the dip in $\sigma_{xx}(0)$ near the charge neutral point in Fig. 2D thus agree with the theoretical prediction for a QAH system with residual dissipative channels. These channels are expected to vanish completely at zero temperature (*11, 24*).

To confirm the QAH effect observed in Fig. 2, we apply a magnetic field, aiming to localize all possible dissipative states in the sample. Figure 3A and 3B display the magnetic field dependence of $\rho_{yx}$ and $\rho_{xx}$ of the same sample as in Fig. 2, respectively. Except for the large MR at $H_c$, increasing the field further suppresses $\rho_{xx}$ towards zero. Above 10 T, $\rho_{xx}$ vanishes completely, corresponding to a perfect QH state. It is noteworthy that the increase in $\rho_{xx}$ from zero (above 10 T) to 0.098 $h/e^2$ (at zero field) is very smooth and $\rho_{yx}$ remains at the quantized value $h/e^2$, which indicates that no quantum phase transition occurs, and the sample stays in the same QH phase as the field sweeps from 10 T to zero field. Therefore, the complete quantization above 10 T can only be attributed to the same QAH state at zero field.

The observation of the QAH effect is further supported by the behavior with varying temperatures. In Fig. 4A, we show $V_g$ dependences of $\rho_{yx}(0)$ and $\rho_{xx}(0)$ measured at different temperatures in another sample with the same growth conditions. The $\rho_{yx}(0)$ always exhibits a single maximum, with the peak value considerably suppressed by increasing temperatures, accompanied by the disappearance of the dip in $\rho_{xx}(0)$. The $\sigma_{xx}(0)$ extracted from these measurements (in logarithmical scale, Fig. 4B) exhibits a temperature dependence similar to that in integer QH systems: the drop of $\sigma_{xx}$ is at first rapid, resulting from the freezing of the thermal activation mechanism, and then becomes much slower when the temperature is below 1 K. It can be attributed to variable range hopping (VRH) (*24*), but its exact mechanism remains unknown. Similar to the QHE, zero field $\sigma_{xx}$ is expected to decrease to zero at sufficiently low temperature. In Fig. 4C we plot the relation between $\sigma_{xx}(0)$ and $\delta\sigma_{xy}(0)$ ($\delta\sigma_{xy} = e^2/h - \sigma_{xy}$, which reflects the contribution of dissipative channels). A power law relation $\delta\sigma_{xy} \propto \sigma_{xx}^\alpha$ with $\alpha \sim 1.55$ is obtained. For a ferromagnetic insulator in the VRH regime, the AH conductivity is related to the longitudinal conductivity through $\sigma_{AH} = A\sigma_{xx}^\alpha$ (the power $\alpha$ is ~ 1.6, the prefactor $A$ can be positive or negative depending on materials (*22*)). The above result can thus be qualitatively understood within the VRH framework.

Our results demonstrate the realization of QAH effect in magnetic TIs. Compared to QHE systems, all the samples studied in this work have a rather low mobility (< 1000 cm$^2$/(Vs)). Such robust QAH states not only reflect the topological character of TIs, but also make the QAH systems readily achievable in experiments. Because the realization of QAH effect and dissipationless edge states does not require any magnetic field, the present work paves a path for developing low-power-consumption, topological quantum electronic and spintronic devices.

**Acknowledgments:** The authors would like to thank Xincheng Xie, Changli Yang, and Xiao-Liang Qi for stimulating discussions and Xiao-Cheng Hong and Shi-Yan Li for helps with experiments. We are grateful to National Science Foundation and Ministry of Science and Technology of China, and Knowledge Innovation Program of Chinese Academy of Sciences for financial supports.

**Figure Captions**

**Fig. 1.** Sample structure and properties. (**A**) A schematic drawing depicting the principle of the QAH effect in a TI thin film with ferromagnetism. The magnetization direction (*M*) is indicated by red arrows. The chemical potential of the film can be controlled by a gate voltage applied on the back-side of the dielectric substrate. (**B**) A schematic drawing of the expected chemical potential dependence of zero field $\sigma_{xx}$ ($\sigma_{xx}(0)$, in red) and $\sigma_{xy}$ ($\sigma_{xy}(0)$, in blue) in the QAH effect. (**C**) An optical image of a Hall bar device made from a $Cr_{0.15}(Bi_{0.1}Sb_{0.9})_{1.85}Te_3$ film. The red arrow indicates the current flow direction during the measurements. The light-grey-colored areas are the remained film and the dark-grey-colored areas are bare substrate with the film removed. The black-colored areas are the attached indium electrodes. (**D**) Magnetic field dependence of $\rho_{yx}$ curves of the $Cr_{0.15}(Bi_{0.1}Sb_{0.9})_{1.85}Te_3$ film measured at different temperatures (from 80 K to 1.5 K). The inset shows the temperature dependence of zero field $\rho_{yx}$, which indicates a Curie temperature of 15 K.

**Fig. 2.** The QAH effect measured at 30 mK. (**A**) Magnetic field dependence of $\rho_{yx}$ at different $V_g$s. (**B**) Dependence of $\rho_{yx}(0)$ (empty blue squares) and $\rho_{xx}(0)$ (empty red circles) on $V_g$. (**C**) Magnetic field dependence of $\rho_{xx}$ at different $V_g$s. (**D**) Dependence of $\sigma_{xy}(0)$ (empty blue squares) and $\sigma_{xx}(0)$ (empty red circles) on $V_g$. The vertical purple dash-dotted lines in (B) and (D) indicate the $V_g$ for the charge neutral point ($V_g^0$). A complete set of the data are shown in fig. S3.

**Fig. 3.** The QAH effect under strong magnetic field measured at 30 mK. (**A**) Magnetic field dependence of $\rho_{yx}$ at $V_g^0$. (**B**) Magnetic field dependence of $\rho_{xx}$ at $V_g^0$. The blue and red lines in (A) and (B) indicate the data taken with increasing and decreasing fields, respectively.

**Fig. 4.** Temperature dependence of QAH effect on a different sample. (**A**) $V_g$ dependent $\rho_{yx}(0)$ and $\rho_{xx}(0)$ measured at 90 mK, 400 mK, 1.5 K, and 4 K, respectively. The vertical purple dash-dotted line indicates the $V_g$ for the charge neutral point ($V_g^0$). The variation in the position and width of the $\rho_{yx}(0)$ peak at different temperatures results from the change

in substrate dielectric properties induced by temperature and charging cycles. (**B**) Dependences of logarithmically scaled $\sigma_{xx}(0)$ (empty red circles) and $\delta\sigma_{xy}(0)$ (empty blue squares) at $V_g^0$ on inverse temperature. The dashed lines are guide to the eye. (C) The relation between $\delta\sigma_{xy}(0)$ and $\sigma_{xx}(0)$ at $V_g^0$ in double logarithmic scale. The red dash line indicates the fit with a power law $\delta\sigma_{xy} \propto \sigma_{xx}^{\alpha}$ with $\alpha \sim 1.55$.

Fig. 1

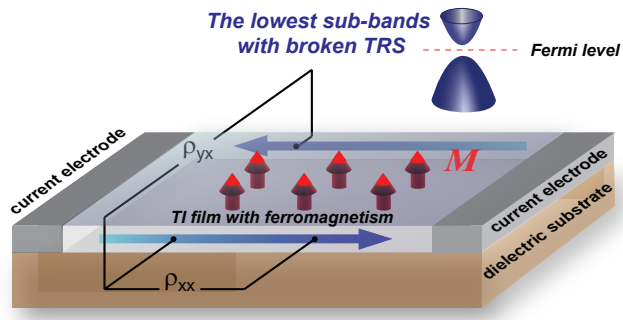
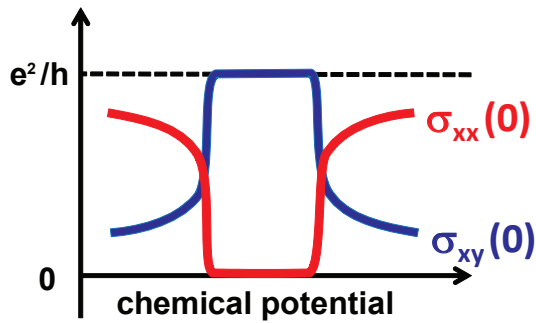
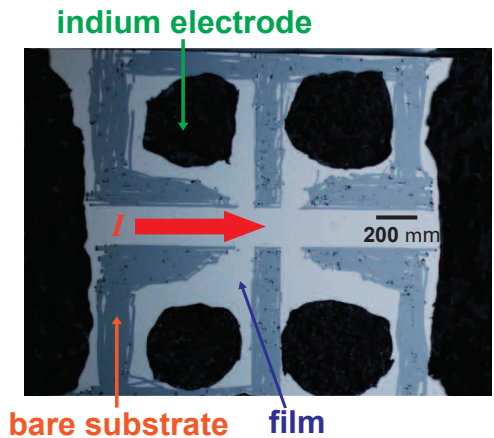
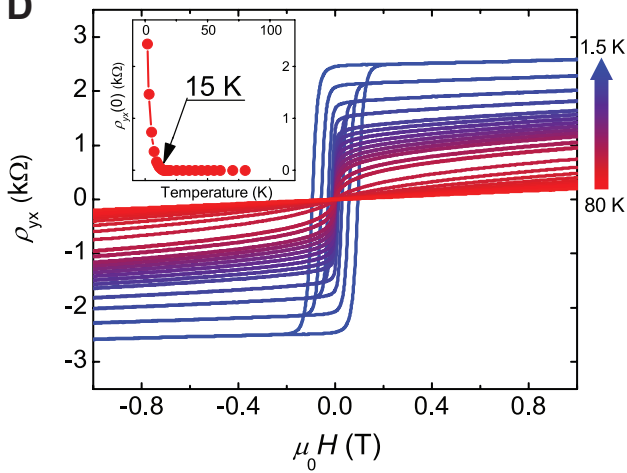

Fig. 2

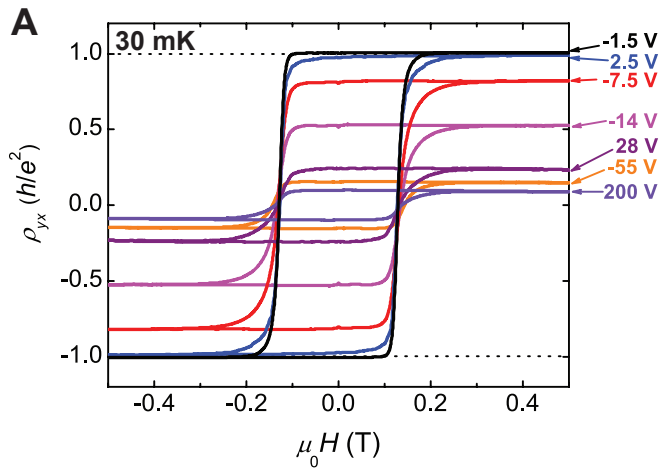
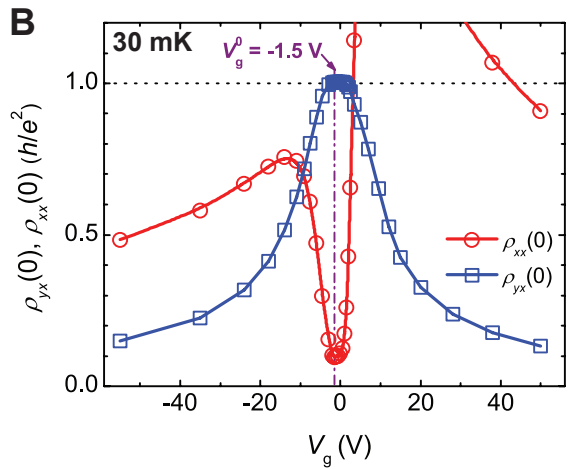
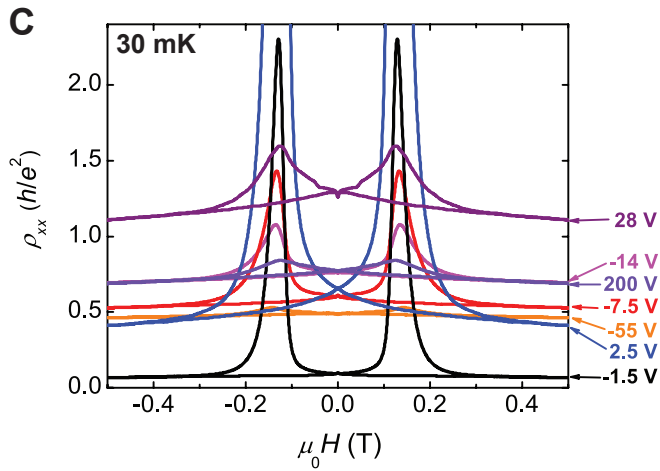
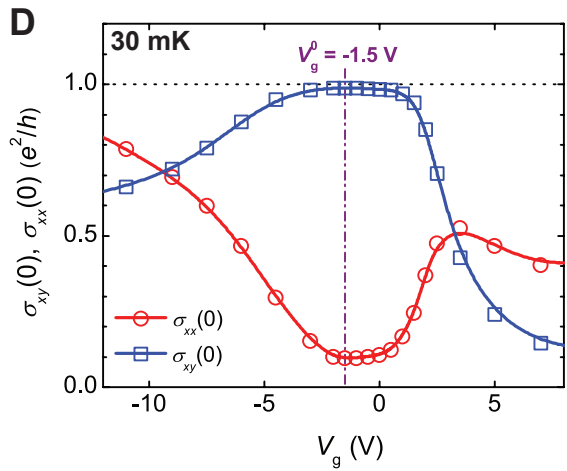

# Fig. 3

**A** 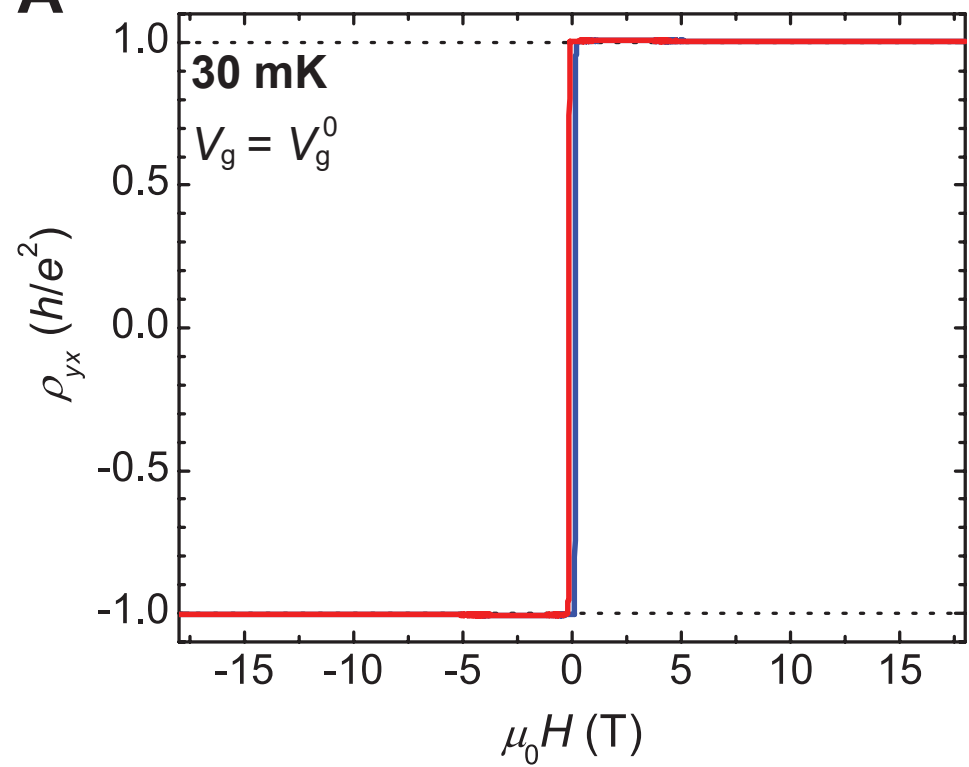

**B** 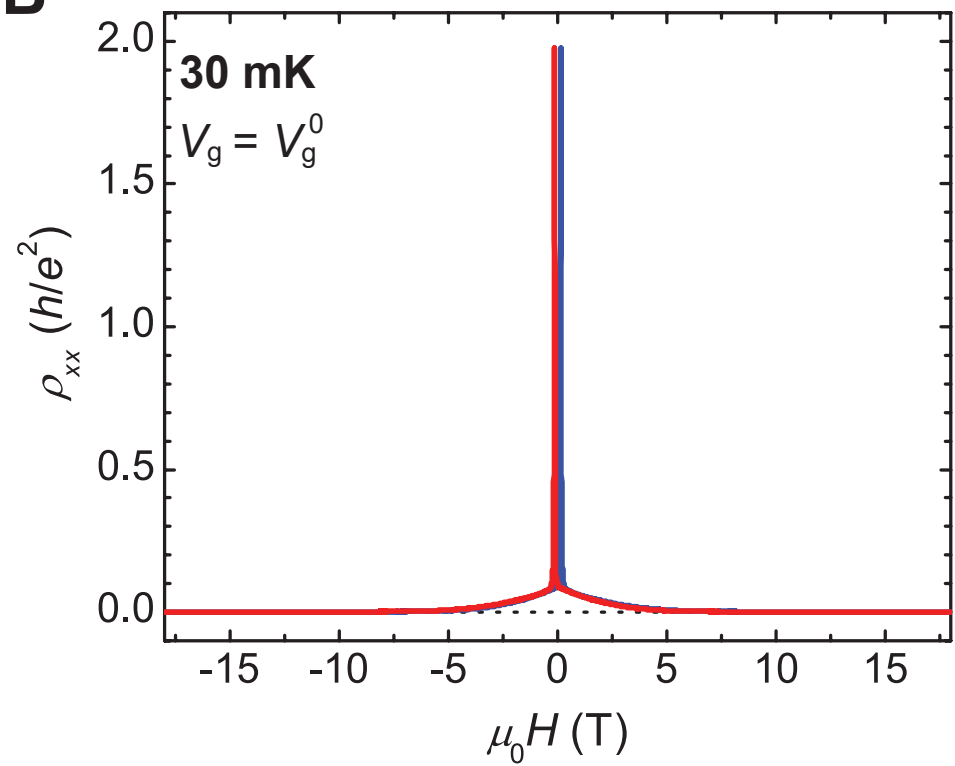

Fig. 4

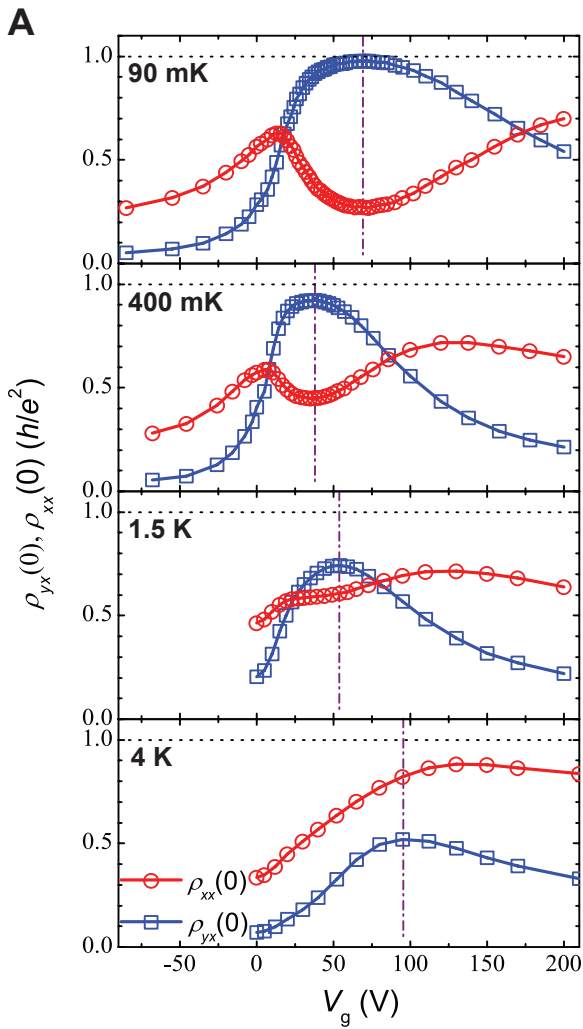
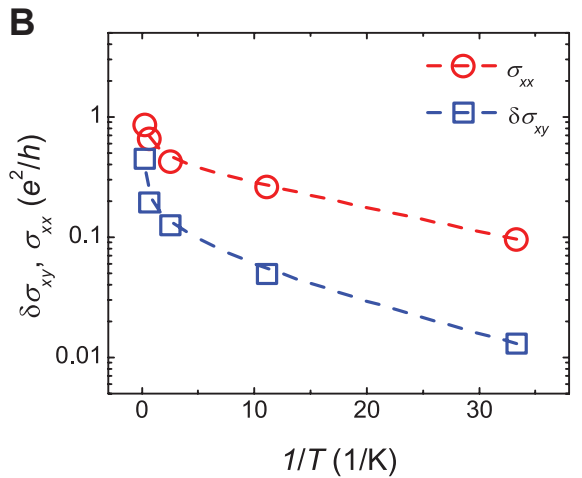
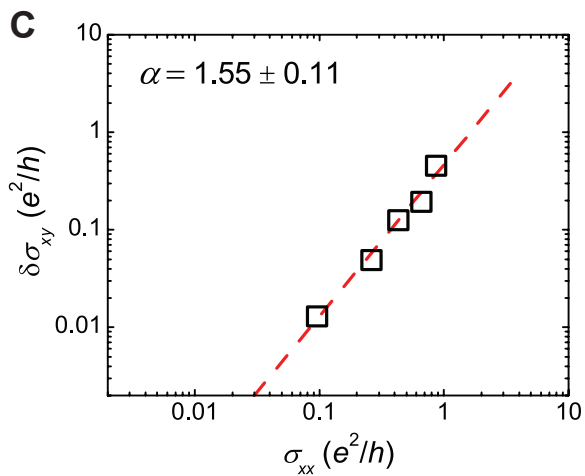